\documentclass[pre]{revtex4}
\usepackage[dvips]{graphics}
\def\frc#1{\frac{1}{#1}}
\def\implies{\;\Longrightarrow\;}
\def\ie{{\it i.e.}}
\def\abs#1{\left|{#1}\right|}
\def\({\left(}\def\){\right)}
\def\[{\left[}\def\]{\right]}
\let\<\le\let\>\ge
\let\al\alpha\let\be\beta
\let\gm\gamma
\let\dl\delta\let\Dl\Delta
\let\ep\epsilon
\let\lm\lambda
\let\om\omega
\def\u#1{\,{\rm #1}}

\def\vmin{v_{\min}}\def\vmax{v_{\max}}
\def\dt{\Dl t}
\def\eqn#1{\begin{equation}#1\end{equation}}
\def\?#1{[\emph{#1}]}
\def\Box{\sqcap\!\!\!\!\sqcup}
\def\V{\widetilde V}\def\tdt{\dl t}
\begin{document}
\title{Numerical analysis of a time-headway bus route model}
\author{Scott A. Hill}
\affiliation{\hbox{}337 Totman Road\\ Lowell MA 01854}
\date{\today}
\begin{abstract}
In this paper, we consider a time-headway model, introduced in
Ref.~\cite{Nagatani}, for buses on a bus route.  By including a
simple no-passing rule, we are able to enumerate and study the unstable
modes of a homogeneous system.  We then discuss the application of the
model to realistic scenarios, showing that the range of reasonable
parameter values is more restricted than one might imagine. We end by
showing that strict stability in a homogeneous bus route requires
careful monitoring by each bus of the bus in front of it, but in many
cases this is unnecessary because the time it takes for the instability
to appear is longer than a bus would normally spend on a route.
\end{abstract}
\maketitle

\section{Introduction}

While there has been much interest in the study of automobile
traffic~\cite{Traffic}, there have been few corresponding studies of
buses~\cite{Buses,Nagatani-old,Nagatani,Nagatani-new}.  The dynamics of
a bus route, while having some similarities with that of automobile
traffic, differs due to the added interaction of buses with passengers
at designated bus stops.  A good reason for studying the dynamics of bus
routes is that they are so often unstable.  Buses are initially spaced
at regular intervals.  However, if one bus is delayed for some reason,
it will find a larger number of passengers waiting for it at subsequent
stops, delaying it further. Meanwhile, the bus following finds fewer
passengers waiting for it, allowing it to go faster until eventually it
meets up with the delayed bus.  Clusters of three, four, or more buses
have been known to form in this manner, resulting in slower service.

In references \cite{Nagatani-old} and \cite{Nagatani}, Nagatani presents
a time-headway model for buses.  Using linear stability analysis, he is
able to determine the range of parameters over which the
homogeneous solution (\ie, with buses spaced evenly apart) is unstable.
 In this paper, we make a more thorough investigation of Nagatani's
model.  We demonstrate the existence of three types of phase diagrams,
in which the behavior of the bus system is divided into four separate
categories.  We conclude with a discussion of how this model
may be applied to real-world situations, and the limitations imposed
by practical considerations.

\section{Model}

\begin{figure}
\begin{center}\includegraphics{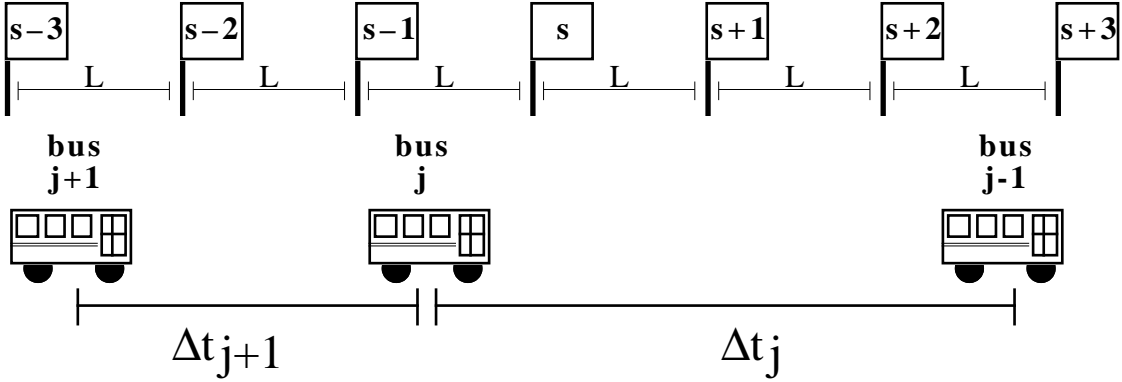}\end{center}
\caption{\label{fig-3buses}Schematic illustration of the model.}
\end{figure}
We consider the following model, introduced in \cite{Nagatani}, of buses
on a bus route (Fig.~\ref{fig-3buses}).  Bus stops are labelled by
$s=1,2,\dots$ where stops $s$ and $s+1$ are a distance $L$ apart.  There
are $J$ buses, $j=1,\dots,J$, which travel from stop to stop, with bus
$j=1$ in the lead and bus $j=J$ in the rear. Every bus visits every
stop, and buses do not pass one another. The time at which bus $j$
arrives at stop $s$ is $t_{j,s}$, which is given by the recursive
relation
\eqn{\label{eq-TimeEvolution}
t_{j,s}-t_{j,s-1}=\lm\gm\,\tdt_{j,s-1}+\frac{L}{V_{j,s-1}},}
where \eqn{\tdt_{j,s}\equiv t_{j,s}-t_{j-1,s}} is the \emph{time-headway}, the
time gap in front of bus $j$ at stop $s$.  
The first term on the right-hand side of Eq.~\ref{eq-TimeEvolution} is
the time it takes for passengers to board the bus at stop $s-1$. The
parameter $\lm$ is the rate at which passengers arrive at a bus stop;
$\lm\,\tdt_{j,s-1}$ is the number of passengers that have arrived at stop
$s-1$ since the previous bus left.  The parameter $\gm$ is the time it
takes each passenger to board the bus, so $\lm\gm\,\tdt_{j,s-1}$ is the
amount of time needed to board all of the passengers.  For convenience,
we introduce the dimensionless parameter $\mu\equiv\lm\gm$, which we
call the \emph{passenger rate}.  For simplicity we ignore the passengers
getting off of the bus; we will assume that it takes less time for the
passengers to get off than it does to get on and pay the fare.

The second term in Eq.~\ref{eq-TimeEvolution} is the time it takes for
bus $j$ to travel from stop $s-1$ to stop $s$, where $V_{j,s-1}$ is the
average velocity of the bus between stops. If this velocity is constant,
then the tendency for buses to bunch together, as described in the
introduction, has no counterweight, and a steady flow of buses will
always be unstable (unless there are no passengers). It is reasonable to
assume, however, that a bus driver will try to prevent bunching by
slowing down when the gap between his bus and the next is too small. 
One can model this by writing the average speed $V_{j,s}$ as a function
$\V(\tdt_{j,s})$ of the gap between his bus and the bus in front of him:
\eqn{\label{eq-TildeVelocityDefinition}
\V(\tdt)=\vmin+(\vmax-\vmin)\frac{\tanh\om(\tdt-t_c)+\tanh\om t_c}{1+\tanh\om t_c}}
The hyperbolic tangent factor acts as a spread-out step function,
centered at $t_c$ with a width proportional to $1/\om$. The parameter
$\vmax$ is the speed a free bus (\ie one that is alone on the route)
would travel.  On the other hand, $\vmin$ is the speed a bus travels if
it has completely caught up with the bus in front of it.  For example,
if $\vmin=0$ then a bus which has caught up with the bus in front of it
will stop and wait until its predecessor has cleared the next stop
before proceeding.

In what follows, it is convenient to work with the time headways
$\tdt_{j,s}$, rather than the arrival times $t_{j,s}$. It is also
convenient to rewrite our expressions in terms of dimensionless
quantities. In doing so, we find that there are four significant
parameters, not including initial conditions.  The first such parameter
is the passenger rate $\mu$. The other three are $\al=L\om/\vmax$,
$\be=\vmin/\vmax$, and $\ep=1-\tanh \om t_c$ (which will typically be
small). We will also consider the dimensionless variable
$\dt_{j,s}=\om\,\tdt_{j,s}$ and the dimensionless velocity function
$V(\dt)=\V(\dt)/\vmax$.  Our evolution equation
Eq.~\ref{eq-TimeEvolution} now reads
\eqn{\dt_{j,s}-\dt_{j,s-1}=\al\[\frc{V(\dt_{j,s-1})}-\frc{V(\dt_{j-1,s-1})}\]+\mu[\dt_{j,s-1}-\dt_{j-1,s-1}],\label{eq-HeadwayEvolution}}
where
\eqn{V(\dt)=\be+\frac{(1-\be)\ep\tanh\dt}{1-(1-\ep)\tanh\dt}=\frac{\be(1-\tanh\dt)+\ep \tanh\dt}{(1-\tanh\dt)+\ep \tanh\dt}.\label{eq-VelocityDefinition}}

\section{Stability Analysis}
We are interested in the stability of a homogeneous flow of buses, with
$\dt_{j,s}=\dt_{j,0}=\dt_0$. One can easily verify that this is a
solution to Eq.~\ref{eq-HeadwayEvolution}. Starting with a small
perturbation to the initial homogeneous solution:
$\dt_{j,s}=\dt_0+y_{j,s}$, where $y_{j,s}$ is small.  To first order,
Eq.~\ref{eq-HeadwayEvolution} becomes
\eqn{y_{j,s}-y_{j,s-1}=[y_{j+1,s-1}-y_{j,s-1}][F(\dt_0)-\mu],\label{eq-1storder}}
where we have introduced the convenient abbreviation
\eqn{F(\dt_0)\equiv\al\frac{V'(\dt_0)}{V(\dt_0)^2}
	=\frac{\al(1-\be)\ep(1-\tanh^2\dt_0)}{[\be(1-\tanh\dt_0)+\ep \tanh\dt_0]^2}.
	\label{eq-Fdef}}
It can be shown~\cite{Nagatani} that the perturbation is stable if
\eqn{\label{eq-StabilityCondition}F(\dt_0)-1<\mu<F(\dt_0).}

\begin{figure}
\begin{center}\includegraphics{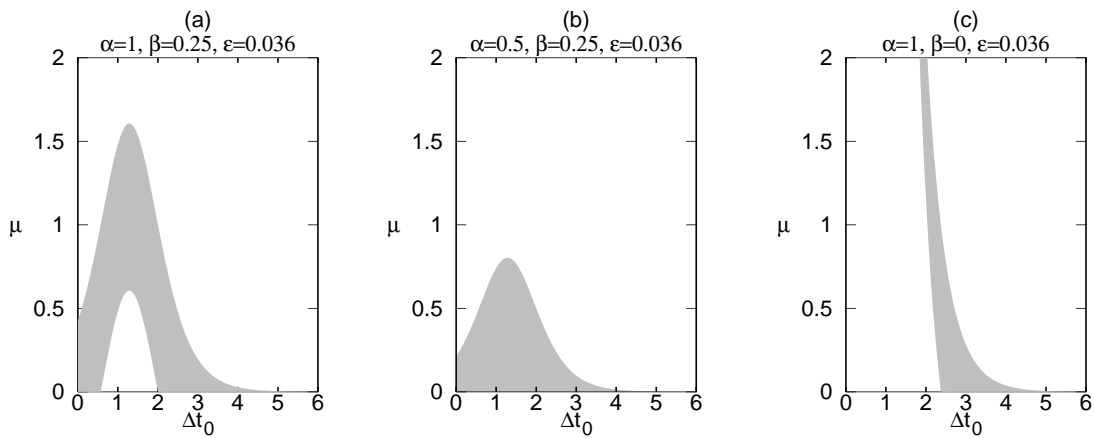}\end{center}
\caption{\label{fig-Fabc}Phase diagrams indicating the regions of
stability of Eq.~\ref{eq-1storder}, for three representative values of
the parameters.  The shaded area is the region which satisfies
Eq.~\ref{eq-StabilityCondition}.}
\end{figure}
From Eq.~\ref{eq-StabilityCondition} we can construct a phase diagram
(Fig.~\ref{fig-Fabc}) for the stability of an initially homogeneous bus
route, based on the initial spacing $\dt_0$ and the passenger rate
$\mu$.  The stable region in phase space is bounded by the curves
$\mu=F(\dt_0)$ and $\mu=F(\dt_0)-1$.  Because of the added constraint
that $\mu\>0$, there are different phase diagrams depending on whether
$F(\dt_0)-1$ is ever positive (Fig.~\ref{fig-Fabc}a) or not
(Fig.~\ref{fig-Fabc}b).  The curve $F(\dt)$ has a maximum value of
\eqn{\al\frac{(1-\be)}{2\be-\ep} \hbox{ at
}x=1-\frac{\ep}{\be}.\label{eq-Fmax},} so the phase diagram resembles
Fig.~\ref{fig-Fabc}a whenever
\eqn{\al>\frac{2\be-\ep}{1-\be}\approx\frac{2\be}{1-\be}.} A third phase
diagram, Fig.~\ref{fig-Fabc}c, arises when $\vmin=0$, as it is in figure
3 of Ref.~\cite{Nagatani} (although apparently not in figure 8 of the
same reference, which may account for the discrepancy between those two
phase diagrams.)

\section{Simulation}
To study the ways in which the system becomes unstable, we evaluate
Eq.~\ref{eq-HeadwayEvolution} iteratively in $s$.  Our initial condition
is \eqn{\dt_{j,0}= \dt_0+0.1r_j,} where $r_j$ are random numbers chosen
between $-1$ and $1$. For each combination of initial headway $\dt_0$
and passenger rate $\mu$, we run the simulation until either a) we reach
stop $s=5000$, or b) one or more of the bus headways exceeds $\dt=1000$ (in
which case the system has become unphysical).

In this paper we consider two different boundary conditions. The first
is periodic in the bus number $j$; so for example
$\dt_{1,s}=t_{1,s}-t_{J,s}$.  This is convenient numerically, and it
creates translational symmetry, but it is hard to construct a physical
model which begins with this characteristic.  We also consider a fixed
boundary condition, where $\dt_{1,s}=\dt_0$. Since the velocity of a bus
depends entirely on $\dt$, this corresponds to a scenario where the
initial bus ($j=1$) moves at a constant speed $V(\dt_0)$.\footnote{ A
more realistic boundary condition may involve ramping up the passenger
rate $\mu$ over time, as occurs during the course of a normal day.}

The structure of the model requires that buses not pass one another;
however, there is nothing in Eq.~\ref{eq-HeadwayEvolution} to prevent
the headways $\dt$ from becoming negative.  To fix this, we add to our
simulation the rule that any $\dt_{j,s}<0$ is replaced by $\dt_{j,s}=0$.
This corresponds to a situation where drivers are forbidden (or unable
due to road conditions) to pass one another.\footnote{An alternative solution
which we do not consider here is to allow buses to pass one another. 
This could conceivably be done by replacing all $\dt_j<0$ with
$\abs{\dt_j}$, effectively swapping the labels of buses that pass one
another.  We have not considered whether this would work in practice,
however.}

\begin{figure}
\begin{center}\includegraphics{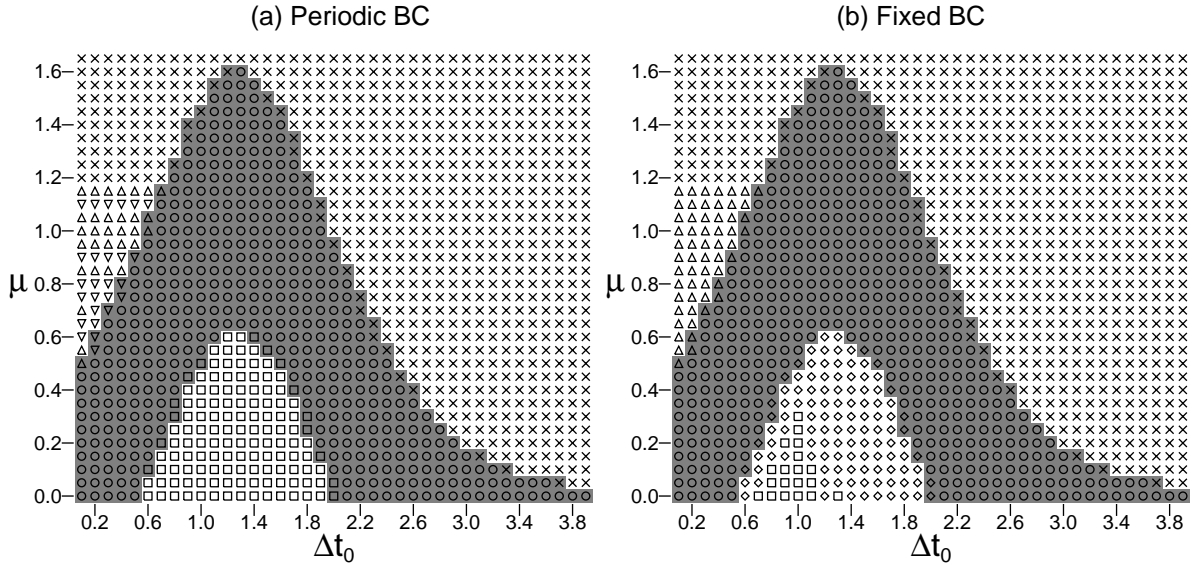}\end{center}
\caption{\label{fig-phasediagrams}Phase diagram for bus systems with (a)
periodic and (b) fixed boundary conditions, where $\al=1$, $\be=1/4$,
and $\ep=0.036$.  The horizontal axis is the initial time headway
$\dt_0$, while the vertical axis is the passenger rate $\mu$.  Stable
runs (\S~\ref{ssec-stable}) are marked by circles ($\bigcirc$) and
exploding runs (\S~\ref{ssec-explosive}) by exes ($\times$). Oscillatory
solutions (\S~\ref{ssec-oscillatory}) are marked by squares ($\Box$);
diamonds ($\diamond$) mark runs which started like oscillatory solutions
 but ended up flat. Slowed solutions (\S~\ref{ssec-slowed}) with
clusters are marked by upward-pointing triangles ($\bigtriangleup$) and
slowed solutions without clusters by downward-pointing triangles
($\bigtriangledown$). The grey shading shows the region where
$F(\dt_0)-1<\mu<F(\dt_0)$.}
\end{figure}

Figure~\ref{fig-phasediagrams} shows the results of our simulation runs
for a typical set of parameters ($\al=1$, $\be=1/4$,
$\ep=\hbox{$1-\tanh2$}=0.036$), using both types of boundary conditions.  In both
cases, the phase space is divided into four regions, corresponding to
four types of runs.

\subsection{Stable Runs}\label{ssec-stable}
\begin{figure}[hpt]
\begin{center}\includegraphics{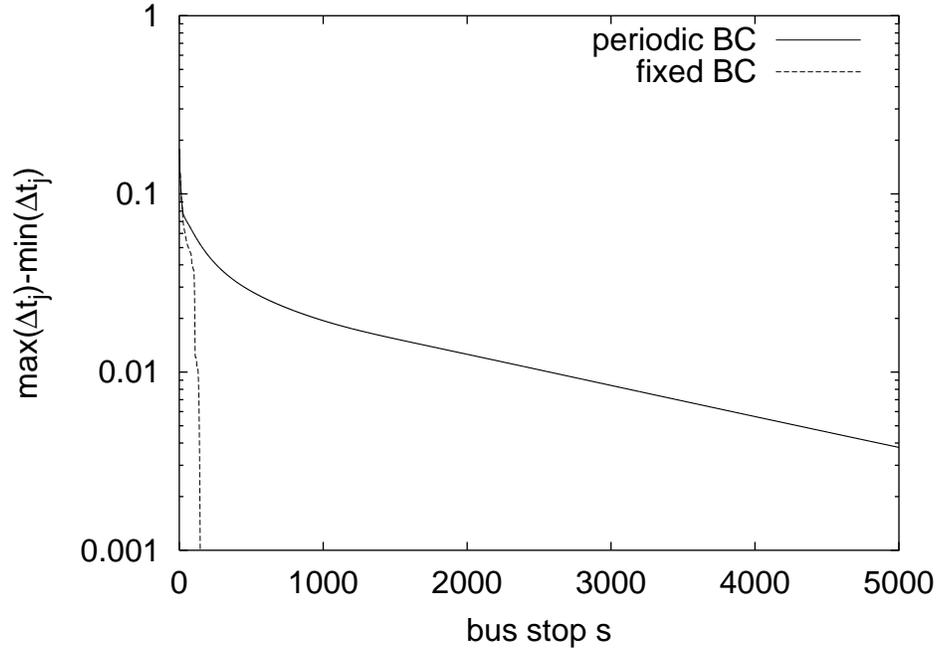}\end{center}
\caption{\label{fig-StableExample} A plot of $\max(\dt_j)-\min(\dt_j)$
versus iteration step $s$ shows that the periodic system ($\mu=0.8$,
$\dt_0=1.5$) is converging exponentially to the homogeneous solution. 
The same system with the fixed boundary condition converges much more
quickly.}
\end{figure}

Most of the runs within the stable region, as defined by
Eq.~\ref{eq-StabilityCondition}, remain homogeneous. In the periodic
case, the initial fluctuations in $\dt_j$ settle into a small precessing
sinusoidal perturbation which decays exponentially with time
(Fig.~\ref{fig-StableExample}).   In the fixed case, the system quickly
locks onto the constant solution $\dt=\dt_0$ with no fluctuations.

\subsection{Explosive Runs}\label{ssec-explosive}
\begin{figure}[hpt]
\begin{center}\includegraphics{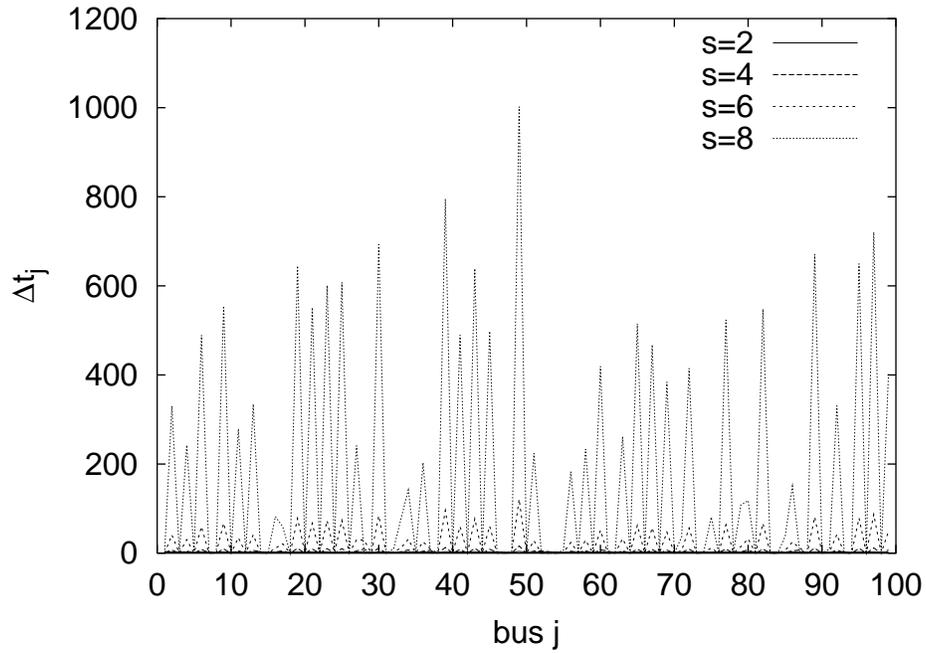}\end{center}
\caption{\label{fig-expev}An extreme example of an explosive run, with
$\mu=1.9$ and $\dt_0=2.5$. By stop $s=8$ there are buses which are
already 1000 time units apart (where one time unit is the time it takes
for a free bus to travel from one stop to the next).}
\end{figure}

Most of the runs lying above the stable region quickly develop an
unphysical instability. This takes the form seen in
Fig.~\ref{fig-expev}, independent of boundary condition: those headways
lying above the mean increase exponentially, while those lying below
decrease steadily until they reach zero.  An observer stationed at a
stop far down the line will see clusters of buses arriving after long
waits; far enough down the line, these waits become astronomical, which is absurd.
Clearly this model is insufficient to deal with these runs at long times.

\subsection{Slowed Runs}\label{ssec-slowed}
\begin{figure}[hpt]
\begin{center}\includegraphics{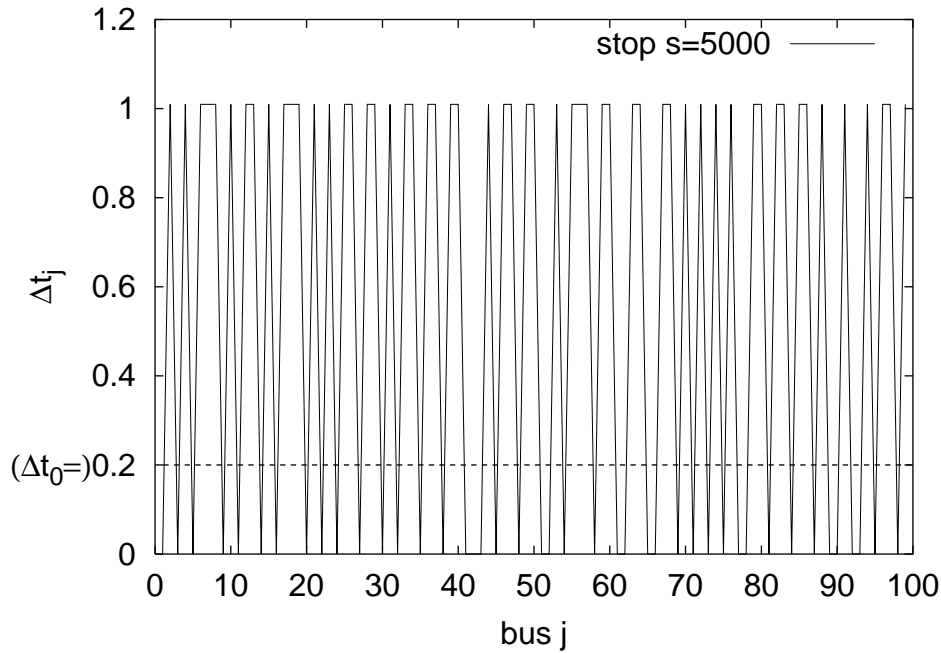}\end{center}
\caption{\label{fig-slowres}An example of a slowed run, where $\mu=0.95$ and $\dt_0=0.2$.}
\end{figure}

To the left of the stable region are runs which develop an alternative
stable solution, as seen in Fig.~\ref{fig-slowres}.  In the case of the
fixed boundary condition, these runs have two things in common.  The
first, indicated by the vanishing of one or more headways, is the
appearance of clusters: two or more buses which travel along as a single
unit. The second is that the units, whether single buses or clusters,
are homogeneously spaced, but with a spacing that is larger than the
initial spacing $\dt_0$.  It should be pointed out that the solutions
shown here are stationary; the clusters and spacings, after they form,
do not change.

\begin{figure}
\begin{center}\includegraphics{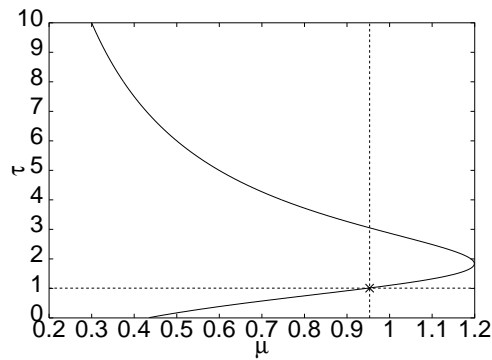}\end{center}
\caption{\label{fig-dt1sol}The solution to Eq.~\ref{eq-Slowed} for the
set of parameters $\al=1$, $\be=1/4$, $\ep=0.036$.  The $\times$ marks
the solution corresponding to Fig.~\ref{fig-slowres}.  In simulation,
the spacing $\tau$ between slowed buses always comes from the lower
branch of the curve.}
\end{figure}
In analytic terms, these states are of the form
$\dt_{j,s}=\dt_j=\tau\,r_j$, where $r_j$ is either $0$ or $1$, and
$\tau>\dt_0$ is a constant. It is straightforward to show that this is a
solution to Eq.~\ref{eq-HeadwayEvolution}:
\eqn{0=\al\[\frc{V(\tau\,r_j)}-\frc{V(\tau\,r_{j-1})}\]+\mu\tau[r_j-r_{j-1}]}
When $r_j=r_{j-1}$, this equation is satisfied trivially.  
Otherwise, the equation takes the form
\eqn{\mu=\frac{\al}{\tau}\(\frc{\be}-\frc{V(\tau)}\).\label{eq-Slowed}}
which we can solve numerically for $\tau$ (Fig.~\ref{fig-dt1sol}). For a
given passenger rate $\mu$, these spacings $\tau$  correspond precisely
with those seen in simulation.  Furthermore, for high enough passenger
rates--- $\mu>1.199$ for this set of parameters--- Eq.~\ref{eq-Slowed}
has no real solutions, which explains the cut-off in
Fig.~\ref{fig-phasediagrams} between the slowed and explosive regimes.

In the case of the periodic boundary condition there are cases where the
clusters eventually break up, leaving a system of buses which are
equally spaced, but with the larger spacing predicted by
Eq.~\ref{eq-Slowed}.  These runs are marked by downward-pointing
triangles ($\bigtriangledown$) in Fig.~\ref{fig-phasediagrams}.

\subsection{Oscillatory Runs}\label{ssec-oscillatory}

\begin{figure}
\begin{center}\includegraphics{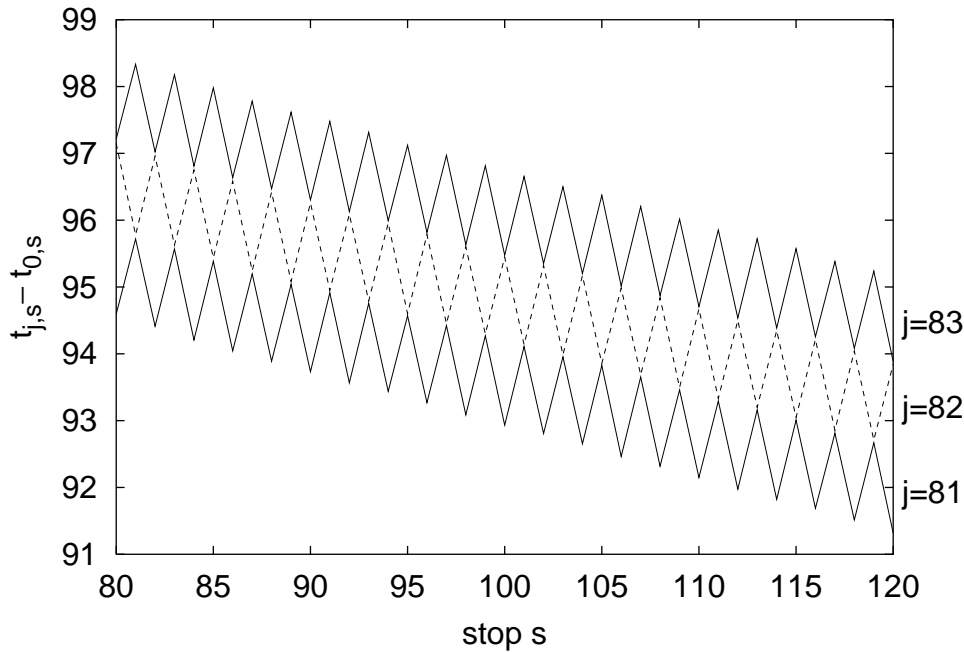}\end{center}
\caption{\label{fig-UnderdampedExample}An example of the oscillatory
condition, with $\mu=0.1$, $\dt_0=1.0$, and fixed boundary conditions. 
This plot shows how long each of three consecutive buses arrive at a
stop $s$ after the initial bus $j=1$ arrived.  Notice that the middle
bus is bunched first with the bus preceding it, then the bus following,
and so forth.}
\end{figure}
\begin{figure}
\begin{center}\includegraphics{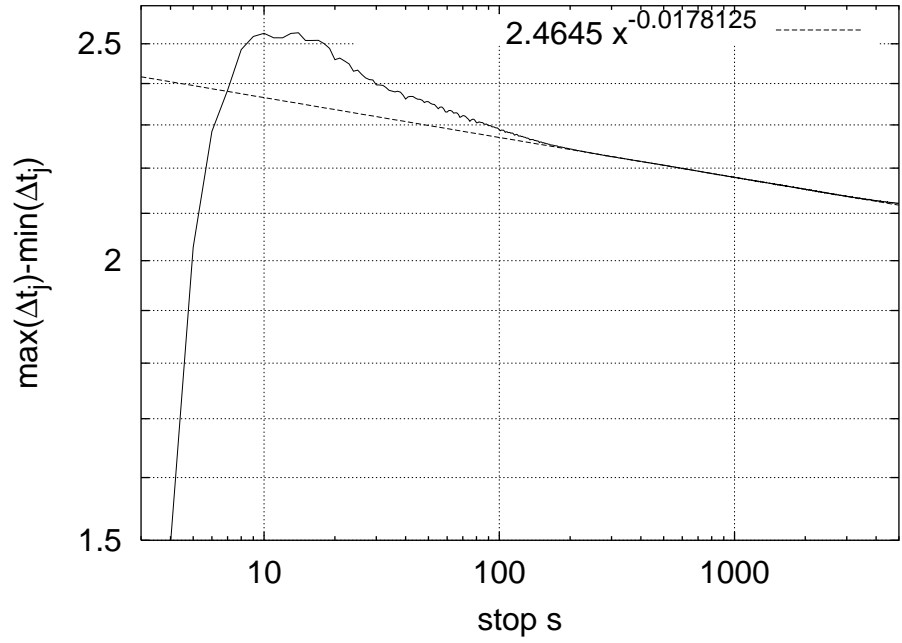}\end{center}
\caption{\label{fig-uddecay}The decay of an oscillatory run with
periodic boundary conditions, $\mu=0.2$, $\dt_0=1.2$.  Both axes are
logarithmic. The power-law decay is too small to have an appreciable
effect on the behavior of the buses.}
\end{figure}
In the case of a run lying below the stability region in phase space,
the first term in Eq.~\ref{eq-HeadwayEvolution}, which is meant to
resist the tendency for buses to cluster, becomes too large.  This leads
to overreaction, so that two buses which arrive too close together at
one stop are too far apart at the next.  The resulting behavior may be
compared to a system of underdamped oscillators.
Figure~\ref{fig-UnderdampedExample} shows the resulting behavior. For
the periodic boundary condition, these oscillations decay as a
power law(Fig.~\ref{fig-uddecay}), but at so slow a rate as to be
practically permanent.
\begin{figure}
\begin{center}\includegraphics{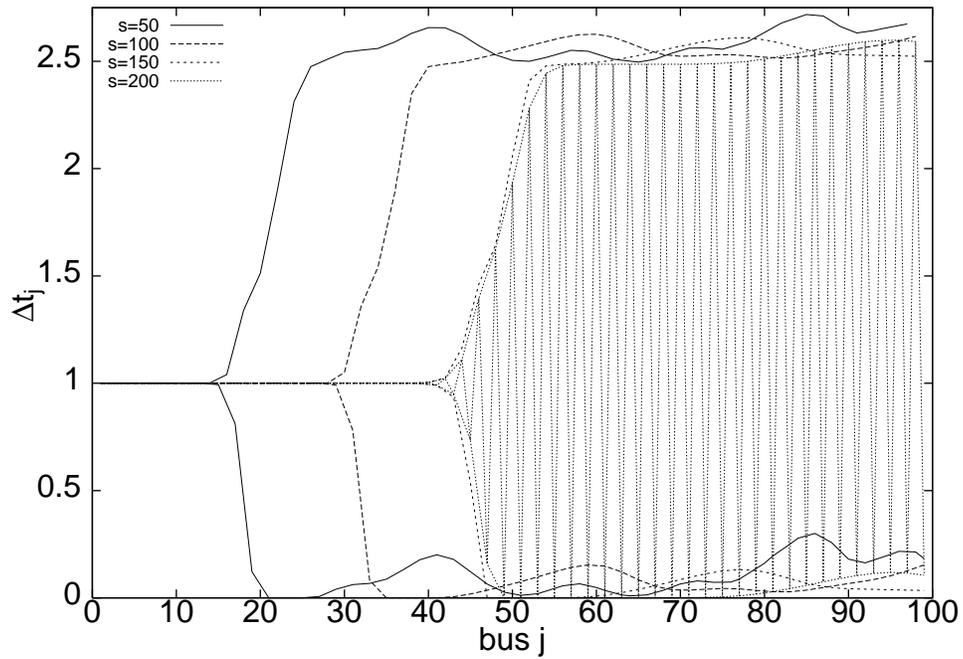}\end{center}
\caption{\label{fig-udev}The headways of an oscillatory system with fixed boundary
conditions, $\mu=0.1$, $\dt_0=1.0$, at four different bus stops.  The bifurcation
of each line indicates the presence of oscillatory behavior, which we show in
full for the $s=200$ case.}
\end{figure}
With the fixed boundary condition, the earliest buses (\ie, those with
the lowest $j$) shed the oscillating behavior after only a few stops,
resuming a homogeneous configuration; with more iterations, more buses
join the homogeneous regime. In some cases, such as in
Fig.~\ref{fig-udev}, the system reaches a steady state with the
oscillations still dominating the later buses.  In other cases, however,
the system becomes completely homogeneous, as the bifurcation point seen
in Fig.~\ref{fig-udev} slips off the right side of the graph. This later
effect may be due to the finite number of buses.

\section{Discussion}

In our simulations we have considered a large range of values for $\mu$
and $\dt_0$. However, these parameters should be limited by a couple of
practical concerns.

The passenger rate $\mu$ is defined as the product of the number of
passengers that arrive at a stop per unit time, and the time it takes a
single person to board the bus.  Put another way, it is the ratio of the
number of people that arrive at a stop to the number of people who can
board the bus in the same amount of time.  If this number is greater
than $1$, then passengers arrive at a stop faster than the bus can take
them on, and the bus should never be able to leave the stop.  Since
$0<\mu<1$, only one of the two inequalities in
Eq.~\ref{eq-StabilityCondition} is meaningful for a given value of the
parameters (including $\dt_0$), depending on whether $F(\dt_0)$ is
larger or smaller than 1.  This suggests that if one wanted to maximize
the area of the stability region in phase space, one would do well to
make sure that the lower stability curve $F(\dt_0)-1$ just grazes zero,
or that $\al(1-\be)=2\be-\ep$ according to Eq.~\ref{eq-Fmax}.

Another practical consideration puts a limit on the value of
$\dt_0$.  Typically, buses are spaced far enough apart so that the first
bus will reach the first stop before the second bus is allowed to leave,
particularly if the stops are spaced fairly close together. This is
described by the inequality
\eqn{\dt_0>\frac{L}{\V(\dt_0)}=\al\frac{(1-\tanh\dt_0)+\ep\tanh\dt_0}{\be(1-\tanh\dt_0)+\ep\tanh\dt_0}>\al.}
For the parameters we have been studying,
\eqn{\dt_0>\frac{1-0.964\tanh\dt_0}{0.25-0.214\tanh\dt_0}\implies
\dt_0>1.82.\label{eq-realisticdt0}} 
This cuts out much of the interesting part of
Fig~\ref{fig-phasediagrams}, including the slowed runs and almost all of
the underdamped solutions. In our discrete model, the basic iteration
step is the bus stop; bus drivers are allowed to change their speed at
the bus stops and nowhere else.  If the buses are several stops apart,
then they have enough time to react to one another.  Otherwise, unusual
situations such as the slowing case or the underdamped case may arise.

Finally, we consider the relationship between $\tdt_0$ and $t_c$,
which is how close a bus will come to the bus in front of it without slowing down.
Consider the ratio
\eqn{r=\frac{1-\tanh\dt_0}{\ep}=\frac{1-\tanh\om\tdt_0}{1-\tanh\om t_c}.\label{eq-ratio}}
We can rewrite $F(\dt_0)$ in terms of this ratio:
\eqn{F(\dt_0)=\al(1-\be)\frac{r(2-\ep r)}{[\be r+1-\ep r]^2}.}
If $\ep r=1-\tanh\dt_0\ll 1$ (which it will be if Eq.~\ref{eq-realisticdt0} is valid, since 
$1-\tanh 1.82=0.05$), then 
\eqn{F(\dt_0)\approx \al(1-\be)\frac{2r}{(1+\be r)^2}.}

Now let us consider what values our parameters might take in
real life. A typical urban bus route might have $L=0.5\u{km}$,
$\vmax=50\u{km/hr}=\frac56\u{km/min}$,
$\vmin=15\u{km/hr}=\frac14\u{km/min}$, and $\om=1\u{/min}$; thus
$\al=0.6$ and $\be=0.3$. A bus which runs every 10 minutes might take on
two passengers at every stop, so $\lm=0.2$ people per minute.  If it
takes $\gm=3\u{s}$ for a person to board a bus, then $\mu=\lm\gm=0.01$.

Consider a scenario where bus drivers only react to what they see; that
is, a driver will only slow down if she can see the next bus in front of
her.  It takes a free bus $L/\vmax=0.6$ minutes to travel from one stop
to the next, so a reasonable value for the amount of warning a bus
driver has is on the order of $t_c=1\u{min}$. Typical bus routes have
buses which are spaced much farther apart, perhaps every
$\dt_0=10\u{min}$ or more.  In this scenario, $r=(1-\tanh 10)/(1-\tanh
1)=10^{-8}$, so $F(\dt_0)\approx 10^{-8}$.  Since $\mu=10^{-2}$, the
stability condition Eq.~\ref{eq-StabilityCondition} is very clearly
violated. For $F(\dt_0)$ to reach a high enough value to create
stability, we need in general for the ratio $r$ to be closer to $1$.
$F(r)$ takes its maximum value when $r=1/(2\be)$, in which case
$F(r)=1/(8\be)=0.4$, which is easily larger than $\mu$ in this example.
Notice that, for values of $t_c$ and $\tdt_0$ greater than $1\u{min}$,
$r\approx e^{2(t_c-\tdt_0)}$, so for each minute's difference between
$t_c$ and $\tdt_0$, $r$ is increased or decreased by a factor of ten.  
It would seem that to maintain a stable homogeneous bus route,
drivers must be reacting to the bus in front of them even from the
very beginning, and can only ignore the leading bus if they have gotten
far enough behind (in which case, of course, the proper solution is to go
as quickly as possible).

\begin{figure}
\begin{center}\includegraphics{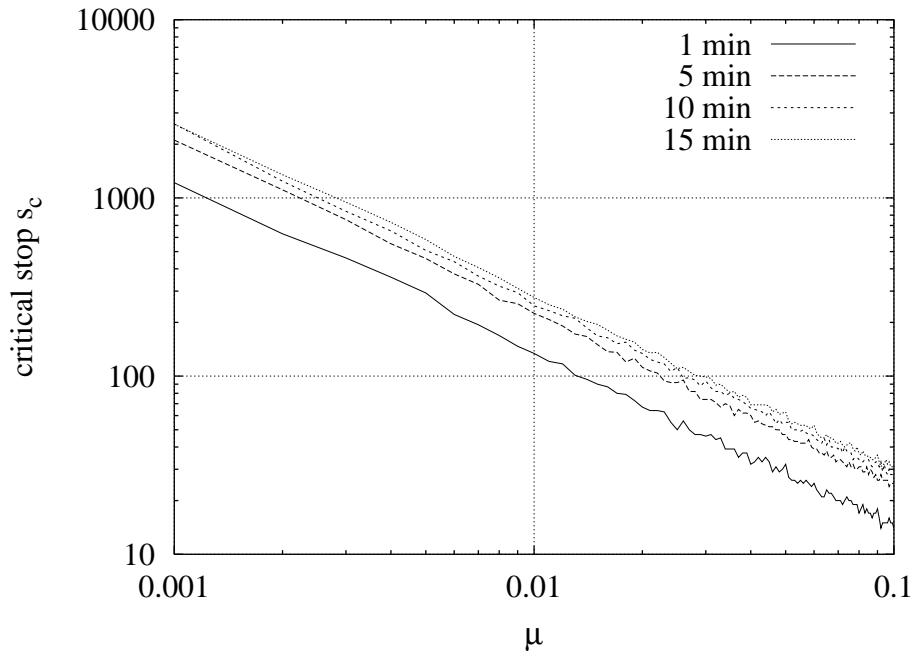}\end{center}
\caption{\label{fig-timetoblowup}This shows the number of stops, for
a given passenger rate $\mu$, that the bus route will run before seeing
a 1-minute, 5-minute, 10-minute, or 15-minute deviation in the initial
homogeneous state.  The parameters used in this simulation are
$\al=0.6$, $\be=0.3$, $\ep=1-\tanh 1$, and $\tdt_0=60\u{min}$.  A power-law
fit to all four lines shows that they all go as $\mu^{-0.965\pm0.005}$.}
\end{figure} 
This is quite a stringent requirement for stability, and explains why it
is so common to see clusters of buses in large cities. It does not seem
likely, however, that this would be the case for less frequent bus
service, such as when buses run once per hour.  A driver
on such a route does not typically keep track of what the previous bus was doing an
hour ago, and yet one does not see clustering behavior
on these low-frequency routes.  The reason for this is that the
instabilities predicted by Eq.~\ref{eq-StabilityCondition} may take a
long time to become noticeable, and normal bus routes tend to have
a limited number of stops. Figure~\ref{fig-timetoblowup} shows the number
of stops that a bus route has to cover before seeing a noticeable
deviation in the initial homogeneous state.  If the passenger rate is
$\mu=0.01$ as suggested in the urban case above, then the route would
have to have 130 stops to show a 1 minute deviation from the homogeneous
state, and 225 stops to show a 5 minute deviation. If $L=0.5\u{km}$,
these correspond to $65\u{km}$ and $112\u{km}$, longer than your average
bus route.  The situation is even better when you consider that a
suburban or rural bus route might have, not 1 person for every 5
minutes, but maybe 1 person every 25 minutes, so that $\mu=0.002$, and
we can start having bus routes with $600$ or $1000$ stops before the
instability becomes noticeable. This is not to say that smaller routes
remain perfectly on time, of course; just that the delays are unlikely
to be due to the need to pick up extra passengers. Since buses will
typically complete the route only to turn around and do it again, one
might consider an entire day's run to be a single route, in which case
instabilities may creep in late in the day. However, the introduction of
a bus terminal, where buses wait until a specific time to leave for
their next trip through the route, would have to be accounted for in
this case.

In this paper, we have considered the bus route model proposed in
Ref.~\cite{Nagatani}. We have added a simple way to deal with negative
time-headways (by replacing all negative $\dt_j$'s with zeroes), and by
doing so have been able to study the unstable modes of a homogeneous
system of buses.  We show that there are in fact three different phase
diagrams (Fig~\ref{fig-Fabc}) for this system, depending on our choice
of parameters, and that in addition to the stable homogeneous state,
there are three unstable modes which the system can fall into: the
explosive mode, the slowed mode, and the underdamped mode.  We then
considered the application of this model to real-life bus routes.  We
have shown that the passenger rate $\mu$ and the initial spacing between
buses $\dt_0$ are greatly restricted by practical considerations, and
that to guarantee stability one needs to have bus drivers who are
constantly tracking the bus in front of them, even when that bus is at
its normal distance.  Fortunately, this is only necessary for bus routes
with very many stops; with fewer stops to make, a bus may be able to
complete the route before any instabilities can become noticeable.

The author thanks Dr. Raghuveer Parthasarathy for indirectly and
unknowingly encouraging this work; and Professor Gene Mazenko for the
computing power needed to run the simulations.  This work was supported
by the Materials Research Science and Engineering Center through Grant
No. NSF DMR 9808595.

\end{document}